\documentclass[psfig]{jltp}

\usepackage{psfig}

\renewcommand{\Re}{{\rm Re}}
\newcommand{\Am}{\AA$^{-1}$}

\newcommand{\Qv}{{\bf Q}}

\newcommand{\beq}{\begin{equation}}
\newcommand{\eeq}{\end{equation}}
\newcommand{\beqa}{\begin{eqnarray}}
\newcommand{\eeqa}{\end{eqnarray}}
\newcommand{\refe}[1]{(\ref{#1})}
\newcommand{\betat}{\tilde\beta}
\newcommand{\Gc}{{\cal G}}
\newcommand{\rem}[1]{}

\title{
Theory and Data Analysis for the High Momentum End 
of $^4$He spectrum
}

\author{F. Pistolesi}

\address{
Institut Laue-Langevin,
B.P. 156, F-38042 Grenoble Cedex 9, France}

\runninghead{F. Pistolesi}{
High Momentum End of $^4$He Spectrum
}

\begin{document}

\maketitle

\begin{abstract}
The hybridization of the single-excitation branch 
with the two-excitation continuum is reconsidered from the 
theoretical point of view by including the effect of 
the interference term between one and two excitations.
The phenomenological theory presented is used to 
reproduce the experimental data over a wide region of 
momentum (2.3-3.6 \Am) and energy (0-12 meV). 
It is thus possible to extract the final part of $^4$He 
spectrum with higher accuracy.
It is found that data agree with a dispersion relation always 
below twice the roton energy. 
This is consistent with the negative 
value of the roton-roton interaction found.

\end{abstract}

\section*{INTRODUCTION}

Excitations in superfluid $^4$He 
have been widely studied
in the last decades.\cite{Griffin} 
However, the nature of the single particle 
spectrum termination is still unclear.  
Pitaevskii a long time ago\cite{Pitaevskii59}
predicted a termination of the spectrum due to a 
decay into pairs of rotons. 
Qualitatively the theory predicts that 
the low energy pole is repelled by the continuum, so that 
the spectrum flattens out for large $Q$ towards 2$\Delta$ losing 
spectral weight ($\Delta$ is the roton energy). 
At the same time, 
a damped excitation for $\omega>2\Delta$
appears and shifts to higher energies.
More precisely, the end of the spectrum is determined by the sign of the 
roton-roton ($V_4$) interaction. 
If this is positive, the spectrum
terminates non analytically {\em at} $2\Delta$ and {\em at} a 
definite momentum $Q_c$.
On the contrary, if $V_4<0$ the spectrum should 
go smoothly to the two-roton bound state energy slightly below
$2\Delta$. In this second case a sharp peak should survive up to
large values of momentum, even if its spectral weight should 
decrease very rapidly. (This second scenario was proposed by
Zawadowski-Ruvalds-Solana\cite{RZ} and Pitaevskii.\cite{Pit70})  
In both cases no damping of the excitations 
is possible below $2\Delta$, nor the {\em sharp excitation} can
exceed $2\Delta$.

Neutron scattering experiments suggest that the decay of excitations 
into {\em pairs of rotons} does actually take place for 
momentum $Q > 2.6$~\AA$^{-1}$.\cite{CowleyWoods}
Despite the good {\em qualitative} agreement between theory 
and experiment the sign of $V_4$ is still undetermined
and the possibility that the excitation energy exceeds $2\Delta$
seems suggested by the data.
Direct fitting of Pitaevskii-ZRS theory to data\cite{Smithetal} 
indicates that $V_4>0$, but  
experimental finding of a quasiparticle peak 
above $2\Delta$ is not accounted for by 
the theory with reasonable values of the parameters.
As a matter of fact, the position of the quasiparticle energy
was extracted by fitting a Gaussian peak on a background
of constant slope.\cite{Smithetal} 
Furthermore, the experimental finding of a (large) positive $V_4$, 
from the theoretical point of view, is completely inconsistent 
with the observation of a spectrum that does not disappear at 
any finite $Q$.     
Recent experimental investigations by F\aa k and 
coworkers\cite{fak91,fak1} on the temperature 
dependence of the dynamical structure factor $S(\Qv,\omega)$
shows clearly that there is a strong correlation
between the low-energy peak and the high energy continuum 
as $Q$ increases from 2.3 to 3.6 \Am.\cite{fak1} Thus supporting
the hybridization picture. 

In this paper we address these inconsistencies that we believe 
are mainly due to the difficulty of taking properly into account the
contribution of the continuum of excitations 
starting {\em at} $2\Delta$, when, for large $Q$, it becomes more important
than the discrete contribution of the sharp state slightly below
$2\Delta$. 

\section{PITAEVSKII-ZRS THEORIES AND THEIR EXTENSION}

We recall that 
the validity of Pitaevskii and ZRS theories 
is restricted to a small region around 2$\Delta$.  
Indeed Pitaevskii in his original paper\cite{Pitaevskii59}
exploited the logarithmic divergence 
appearing in the two-roton response function
[$
	F_o({\bf Q}, \omega)
	=
	i\,\int {d\omega'\over 2\pi} \int {d^3 \Qv' \over (2\pi)^3}
	\Gc(p-p')\Gc(p')
$,
where $\Gc^{-1}(p)=\omega+i\,0^{+}-\omega({\bf Q})$,
$\omega({\bf Q})$ is the measured spectrum and
$p=(\Qv,\omega)$]
to solve {\em exactly} the many-body equations.
This elegant theory provides an 
explicit expression for the Green function
valid {\em only} in the small energy range where the 
singularity dominates. 
This fact leads to problems in data analysis when 
the bare excitation energy $\omega_o({\bf Q})$ 
reaches values well above $2\Delta$, since the signal around 
$2\Delta$ strongly decreases.
To understand the correlation between the high energy part of the 
spectrum and the one-excitation contribution,
it is thus necessary to extend the validity of the theory
to a wider range of energies 
in order to describe properly the continuum contribution
to $S(\Qv, \omega)$. 
It then becomes crucial to consider the two main channels that contribute 
to the continuum: 
The excitation of a quasiparticle that then decays into a 
pair of quasiparticles and the direct excitation of two quasiparticles.
It is possible to construct a phenomenological theory 
that takes into account this two channels. Details are given 
elsewhere,\cite{mio} here we only report the following 
 simple expression for the density-density response function $\chi$:
\beq
	\chi(p) = 
	{\alpha^2 + 2 \alpha \beta V_3 F(p) 
	+\beta^2 F(p)\,\Gc_o^{-1}(p) 
	\over 
	\Gc_o^{-1}(p) - V_3^2\, F(p)
	}
	\label{eq11}.
\eeq
In this expression (valid at zero temperature) 
$\alpha$ and $\beta$ are the matrix element 
for the excitation of one and two quasiparticles, respectively.
The coupling $V_3$ parametrize the decay amplitude 
and $F(p)$ is given by the sum of all the 
diagrams with two lines joined at the two external legs 
linked at least by two lines.\cite{mio} 
The effect of a $V_4$ interaction is completely 
hidden in $F(p)$, and if we consider only energies near the 
threshold, Pitaevskii-ZRS theory can be used to obtain an explicit 
expression for $F(p)$.
The dynamical structure factor is simply related to $\chi$ by
$S(\Qv,\omega)=-{\rm Im} \chi(\Qv, \omega)$ and Eq.~\refe{eq11}
constitutes the starting point to analyze data.
We also note that Eq.~\refe{eq11} for $\beta=0$ gives the Green 
function of one quasiparticle. The above expression thus 
defines a $\Gc$ and a $\chi$ that share the same poles as 
expected in a Bose condensed system.\cite{Griffin}

\section{DATA ANALYSIS AND RESULTS}

Since the explicit calculation of $F$ is a difficult task and 
in general depends strongly on the detailed structure of the 
vertex functions, we do not try to calculate it microscopically,
but we extract it directly from data by exploiting 
the large (energy and momentum) region of validity of Eq.~\refe{eq11}.
The main idea is to extract a one-dimensional function [$F(\omega)$] 
from a two-dimensional experimental function [$S(\Qv,\omega)$].
As a matter of fact the main $Q$-dependence of Eq.~\refe{eq11}
is through $\omega_o(\Qv)$, since $F(\Qv,\omega)$ is expected 
to depend weakly on $\Qv$ (as is the case for $F_o(\Qv,\omega)$ in this 
momentum region).
It is thus possible to extract $F(\omega)$ and $\omega_o(\Qv)$
by fitting Eq.~\refe{eq11} to the different sets of data with 
different momentum {\em at the same time}. 
This procedure exploits fully the information contained in the 
data because it is sensitive to the correlation among sets with 
different $Q$. 

To extract the complex function $F(\omega)$ from data 
we parametrize its imaginary part with $N$ real numbers 
(15 in the fit presented) in the following way. 
We choose a set of values of $\omega$, say 
$\{\omega_1, \omega_2, \dots, \omega_N\}$ 
with $2\Delta = \omega_1 < \omega_2 <\dots<\omega_N$ reasonably 
spaced and we assign a {\em free} parameter, $a_i$, to each of them.
${\rm Im} F(\omega)$ can then be defined as a cubic spline 
interpolation on such a set. 
The real part can then be easily obtained by the relation
$
\Re F(\omega) 
= -{1/\pi}{\cal P}\int d\omega' {\rm Im} F(\omega)/(\omega-\omega')
$. 
It is also important to reduce the parameters to the minimum
number of independent ones, so we define a dimensionless function 
$f(\omega)=\lambda F(\omega)$,
constrained by the normalization condition 
$
	\int_{2\Delta}^{\omega_N} d\omega {\rm Im} f(\omega)
	=	
	\omega_N-2\Delta
$, and $g_3=V_3/\lambda^{1/2}$, $\betat=\beta /(\lambda^{1/2}\alpha)$.
In this way $\alpha$, $\betat$, $g_3$, and the $N-1$ parameters that define 
$f$ are independent and can be fitted to the data.

\begin{figure}[tcbh]
\centerline{\psfig{file=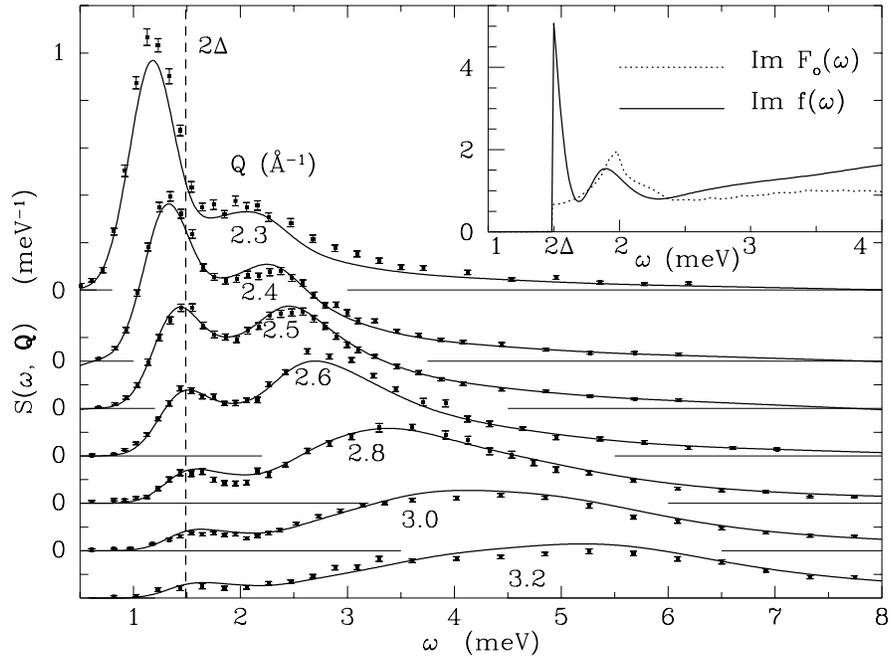,width=5in,angle=-90}}
\caption{
Fit to the data from Ref.~8 
with a parametrized ${\rm Im} f$. In the inset the 
resulting ${\rm Im} f$ is shown compared with ${\rm Im} F_o$ averaged 
over $2.3<Q<3.2$ \AA$^{-1}$ and properly scaled.}
\label{fig2}
\end{figure}

We thus fitted Eq.~\refe{eq11} 
(convolved with the known instrumental resolution) 
to experimental data of Ref.~8 at 1.3 K.
The resulting fit is shown in Fig.~\ref{fig2}. 
The good agreement between theory and experiment is obtained with a  
reduced $\chi^2$ of nearly 4,
thus indicating that even if we are leaving a large freedom in 
$f$ the agreement is significant.
%
% ---------------------------------------------------------------------
% ---------------------------------------------------------------------
%

The new dispersion relation for the undamped excitation is 
reported in Table 1. Note that the value of $2\Delta$ is 
1.484 meV.\cite{Don}
The fitted values of the normalized $a_i$ is also reported in Table 1.
\begin{table}[tcbh]
\begin{center}
\begin{tabular}{|c|c|c|c|}
\hline
Q & $\omega(Q)$ & $Z(Q)$ & $\omega_o(Q)$\\
\Am & meV & & meV\\
\hline
2.3 & 1.174 $\pm$ 0.005 &  0.372  $\pm$  0.019   & 1.42 \\
2.4 & 1.303 $\pm$ 0.003 &  0.234  $\pm$  0.025   & 1.88 \\
2.5 & 1.381 $\pm$ 0.002 &  0.133  $\pm$  0.019   & 2.4 \\
2.6 & 1.420 $\pm$ 0.005 &  0.075  $\pm$  0.008   & 2.8 \\
2.8 & 1.455 $\pm$ 0.010 &  0.031  $\pm$  0.005   & 3.7 \\
3.0 & 1.467 $\pm$ 0.008 &  0.017  $\pm$  0.002   & 4.6 \\
3.2 & 1.473 $\pm$ 0.010 &  0.011  $\pm$  0.001  & 5.3 \\
3.6* & 1.480 $\pm$ 0.010 &  0.003  $\pm$  0.001 & 6.8 \\
\hline
\end{tabular}
\begin{tabular}{|c|c|}
\hline
$\omega_i$ (meV) & $a_i$ \\ 
\hline
 1.484 & 2.7 \\ 
 1.56 & 1.10 \\ 
 1.68 & 0.33 \\ 
 1.75 & 0.45 \\ 
 1.85 & 0.71 \\ 
 2.00 & 0.63 \\ 
 2.15 & 0.44 \\ 
 2.50 & 0.44 \\ 
 3.0 & 0.57 \\ 
 3.5 & 0.67 \\ 
 4 & 0.78 \\ 
 5 & 0.88 \\ 
 6 & 0.97 \\ 
 9 & 1.35 \\ 
12 & 1.09\\
\hline 
\end{tabular}
\end{center}
\caption{Dispersion relation for the quasiparticle spectrum, 
weight ($Z(Q)$) of the pole, and energy of the bare pole $\omega_o$
(left Table). Fitted values for $a_i$ at each 
$\omega_i$ (right Table). The values for $Q=3.6$ \Am are obtained 
from a fit not shown.}
\end{table}

We find that the model can {\em quantitatively} explain 
that the peak position of $S(Q,\omega)$ is 
slightly larger than $2\Delta$ for $Q>2.6$ \AA$^{-1}$.
This originates from a mixing within the instrumental resolution
of the contribution of the sharp peak
at energy $\omega_Q$ slightly smaller than 2$\Delta$ with that 
of the continuum of two
rotons excitations starting at $2\Delta$.
On general grounds the continuum should depend 
strongly on $\omega$ near $2\Delta$, 
as for $\omega>2\Delta$ there are much more states available 
to decay into.
Our procedure exploits the theoretical model for $\chi(p)$ 
to extract both the value of $\omega_Q$ and the continuum contribution. 
In this way we can find the final part of the dispersion 
relation with improved 
accuracy. It turns out that data agree with a dispersion
relation for the excitations always below 2$\Delta$.

Concerning the other parameters of the fit we find
$\alpha^2=1.4$, $\betat=-0.06$ meV$^{-1/2}$ 
and  $g_3=0.8$ meV$^{1/2}$. 
This implies that the direct excitations of two quasiparticles 
by the neutron gives a small but {\em appreciable} 
contribution to $S(\Qv,\omega)$,
particularly at high energy.

The shape of ${\rm Im} f$ found with the fit (see Fig.~1) 
has two main features: A clear peak at $\omega\approx 2$ meV and 
a ``quasi-divergent'' behavior at the threshold. 
The peak is due to the maxon-roton van Hove 
singularity as is clear by comparison with $F_o$.
It is remarkable that although no trace of the 
peak is apparent in any of the experimental plots
the procedure succeeded in finding correctly this information.

The quasi-singularity at threshold can  
be understood as an interaction effect,
namely a signature of the roton-roton attractive interaction.
As a matter of fact, in the 
{\em small region} of energy near the threshold we can 
apply Pitaevski-ZRS theory to evaluate $F(\omega)$.
We verified quantitatively this fact by repeating the fit 
using Pitaevskii-ZRS theory to parametrize $f(\omega)$
and setting a cutoff in energy at $2\Delta$+0.2 meV.
The resulting reduced $\chi^2$ for the fit gives evidence that 
the theory works {\em quantitatively} in this small region.  
We obtained in this way for the interaction parameter 
$V_4 \approx -4.7$ meV \AA$^3$ with a tiny 
bound-state energy of  1.3 $\mu eV$. Note that  
without the cutoff this theory is not able to fit 
data satisfactorily.

In conclusion, we presented a theory for $S(\Qv, \omega)$ that takes into 
account  both one- and two-quasiparticle excitations by the neutron.
The theory reproduces the experimental results over a 
large range of energy and momentum. 
We have thus been able to extract 
the final part of the spectrum dispersion relation in $^4$He and 
to determine its end as an hybridization with the bound state of two
rotons.

\section*{ACKNOWLEDGMENTS}

I am indebted to P.~Nozi\`eres and B.~F\aa k for 
many suggestions. I also thank  B.~F\aa k and J.~Bossy for 
letting me use their data prior to publication.

\end{document}